\documentstyle[aps,prl,psfig,multicol,amstex,epsf]{revtex}

\def\dvg{{\rm div}\,}

\begin{document}

\title{\vbox to 0pt {\vskip -1cm \rlap{\hbox to \textwidth {\rm{\small
SUBMITTED TO PHYS. REV. LETT.\hfill}}}}The Camassa-Holm equations
as a closure model for turbulent channel flow}

\author{Shiyi Chen$^{1}$,Ciprian Foias$^{1,2}$, Darryl D. Holm$^{1}$,
Eric Olson$^{1,2}$, Edriss S. Titi$^{3,4,5}$, Shannon Wynne$^{3,5}$}

\address{${}^{1}$Theoretical Division and Center for Nonlinear Studies,
Los Alamos National Laboratory, Los Alamos, NM 87545\\
${}^{2}$ Department of Mathematics, Indiana University, Bloomington, IN 47405\\
${}^{3}$ Department of Mathematics, University of California, Irvine, CA
92697\\
${}^{4}$ Department of Mechanical and Aerospace Engineering,
University of California, Irvine, CA 92697\\
${}^{5}$ Institute for Geophysics and Planetary Physics,
Los Alamos National Laboratory, Los Alamos, NM 87545 }

\maketitle

\begin{abstract}

We propose the viscous Camassa-Holm equations 
as a closure approximation for the Reynolds-averaged equations
of the incompressible Navier-Stokes fluid. This approximation is tested on
turbulent channel flows with steady mean. Analytical solutions for the mean
velocity and the Reynolds shear stress across the entire channel are
obtained, showing good agreement with experimental measurements and direct
numerical simulations. As Reynolds number varies, these analytical mean
velocity profiles form a family of curves whose {\it envelopes} are shown to
have either power law, or logarithmic behavior, depending on the choice of
drag law.

\end{abstract}

\vskip 4mm

\begin{multicols}{2}

Laminar Poiseuille flow occurs when a fluid in a straight channel, or duct, is
driven by a constant upstream pressure gradient, yielding a parabolic
streamwise velocity profile which is symmetric about the midplane of the
channel. In turbulent states, the mean streamwise velocity profile remains
symmetric, but is flattened in the center because of the increase of the
velocity fluctuation. Although a lot of research has been carried out for
turbulent channel flow, e.g.,~\cite{Reichhardt38,Eckelmann70,Eckelmann74,Wei
Willmarth89,Antonia etal92,Kim-Moin-Moser87}, accurate measurement of
the mean velocity and the Reynolds stress profiles, in particular for
flows at high Reynolds numbers, is still an experimental challenge and
the fundamental understanding of how these profiles
change as functions of Reynolds number is still missing.

In turbulent channel flows it is customary to define a characteristic velocity
$u_*$ and Reynolds number $R_0$ by $u_*=\sqrt{|\tau_0|/\rho}$ and
$R_0=du_*/\nu$, where $\tau_0$ is the boundary shear stress, we take the
density $\rho$ to be unity, $d$ is the channel half-width
and $\nu$ is the molecular viscosity of the fluid. Based on
experimental observation and numerical simulation, a piecewise
expression of the mean velocity across channel has been proposed, for which
the nondimensional mean streamwise velocity, $\phi \equiv U/u_*$, is assumed
to depend on $\eta \equiv u_*z/\nu$ and have three types of behavior depending
on the distance away from the wall boundary, $z$: a viscous sublayer, in which
$\phi\sim\eta$; the von K\'arm\'an-Prandtl logarithmic ``law of the wall,''
in which
$\phi(\eta)={\kappa}^{-1} \mbox{\rm ln} \eta + A$, where $\kappa \simeq
0.41$ and
$A \simeq 5.5$; and a power law region, in which $\phi \sim \eta^p$,
$0<p<1$~\cite{Hinze75}.

In this paper, we propose the viscous Camassa-Holm equations (VCHE) in
(\ref{VCHE-eq}) as a closure approximation for the Reynolds equations. The
analytical solutions of the steady VCHE depend on the three parameters: the
flux Reynolds number $R=d\overline{u}/\nu$ (where $\overline{u}$ is the
streamwise velocity, averaged across the channel); the drag law for wall
friction $D=2\tau_0/\overline{u}^2=2R_0^2/R^2$ as a function of $R$;
 and a shape factor
$c\in(0,1)$ which specifies the flattening of the velocity profile and is
given by the ratio
$u_{max}/\overline{u}=(3-c)/2$ of the maximum streamwise velocity $u_{max}$
(occurring at the center of the channel) to its average $\overline{u}$.
This family of analytical solutions of the steady VCHE provides profiles of
the mean velocity and the Reynolds shear stress depending on the
three parameters  $R$, $D$ and $c$. (Giving these three
parameters is equivalent to giving the Reynolds numbers $R$, $R_0$ and
$R_{max}$, corresponding to the channel flux, boundary stress and velocity at
the midplane, respectively.) The VCHE profiles agree well with data obtained
from turbulent channel flow measurements and simulations across the entire
channel. As the Reynolds number is varied in these solutions, the velocity
profiles $\phi(\eta,R)$ form a family of curves with upper and lower envelopes.
The Blasius $R^{-1/4}$ law~\cite{Townsend} for the drag coefficient of the wall
boundary leads to $\eta^{1/7}$ power law behavior for these envelopes,
while the von K\'arm\'an drag law leads to logarithmic envelopes.

We begin our theoretical treatment by recalling the Reynolds-averaged
Navier-Stokes equations~\cite{Hinze75,Townsend}
\begin{equation} \label{Reynolds-eqs}
	\frac{\partial \langle{\bf u}\rangle}{\partial t}
	+ \langle{\bf u}\rangle \cdot\nabla \langle{\bf u}\rangle
 =\dvg \langle{{\bf T}}\rangle,
\
	\dvg \langle{\bf u}\rangle = 0,
\end{equation}
where $\langle {\bf u} \rangle$ denotes the ensemble average of the
velocity and $\langle{{\bf T}}\rangle=-\langle{p}\rangle{{\bf I}}
- \langle{\bf u}\otimes{\bf u}\rangle +
	\nu(\nabla \langle{\bf u}\rangle +\nabla \langle{\bf u}\rangle
^{\rm T})$.
For turbulent channel flow, the mean velocity is of the form
$\langle{\bf u}\rangle =(\overline{U}(z),0,0)^T$, with
$\langle{p}\rangle = \overline{P}(x,y,z)$ and
the Reynolds equations (\ref{Reynolds-eqs}) reduce to $\dvg \langle{{\bf
T}}\rangle=0$, or
equivalently,
\begin{eqnarray}
\label{Reynolds-eqns}
  &-\nu \overline{U}''+\partial_z \langle wu \rangle
  =-\partial_x\overline{P},& \nonumber \\
  &\partial_z\langle wv \rangle = -\partial_y \overline{P},\quad
  \partial_z \langle w^2 \rangle = -\partial_z \overline{P},&
\end{eqnarray}
where $(u,v,w)^T$ is the fluctuation velocity in the infinite channel
$\{(x,y,z)\in{\mathbb{R}}^3,\,-d\le{z}\le{d}\}$. The (1,3) component of the
averaged
stress tensor $\langle{{\bf T}}\rangle$ is given by
$\langle{T_{13}}\rangle=-\nu
\overline{U}'(z)+\langle{wu}\rangle$. On the boundary, the velocity
components all
vanish and one has the stress condition
\begin{equation} \label{stress-cond}
\mp\tau_0=\langle{T_{13}}\rangle\Big|_{z=\pm d}
=\nu \overline{U}'(z)|_{z=\pm d}\,,
\end{equation}
upon using $\langle{wu}\rangle = 0$ at $z=\pm d$. Hence, the Reynolds
equations imply
$\langle{wv}\rangle(z)\equiv0$ and
$\overline{P} = P_0 -{\tau_0}x/d - \langle{w^2}\rangle(z)$,
with integration constant $P_0$.

The viscous Camassa-Holm equations (VCHE) are
\begin{equation}
\label{VCHE-eq}\hspace{-.4in}
\frac{d{\bf v}}{dt} +	v_j\nabla u^j
     + \nabla\Big(   p
     -\frac{1}{2} |{\bf u}|^2
     -\frac{\alpha^2}{2} |\nabla {\bf u}|^2\Big)
=\nu \Delta {\bf v},
\end{equation}
with $	\nabla\cdot{\bf v} = 0 =\nabla\cdot {\bf u}$, where
${\bf v}=(1-\alpha^2\Delta){\bf u}$,
$d/dt=\partial/\partial t + {\bf u}\cdot\nabla$ is
the material derivative. Also, $\alpha$ is a constant lengthscale,
$ |\nabla {\bf u}|^2 =
    \mbox{Tr}(\nabla {\bf u} \cdot \nabla {\bf u}^T)$
and $\Delta$ is the Laplacian. Equation (\ref{VCHE-eq}) with $\nu=0$ is
derived in~\cite{Holm et al98} by decomposing Lagrangian parcel
trajectories into mean and fluctuating parts, then applying asymptotic
expansions,
Lagrangian means, and assuming isotropy of fluctuations in
Hamilton's principle
for an ideal incompressible fluid. That derivation generalizes a
one-dimensional
integrable dispersive shallow water model studied in~\cite{CH93} to
$n$-dimensions and provides the interpretation of $\alpha$ as the typical
amplitude of
the rapid fluctuations over whose phase the Lagrangian mean is taken in
Hamilton's
principle. Moreover, that derivation makes it clear the solutions of 
VCHE are mean
quantities.

Before comparing VCHE with Reynolds averaged equations, we
rewrite (\ref{VCHE-eq}) in  the equivalent `constitutive' form
\begin{equation}
\label{VCHE-tnsr}\hspace{-.25in}
	\frac{d{\bf u}}{dt}=\dvg {\bf T},\
	{\bf T}=-p{\bf I}+2\nu(1-\alpha^2\Delta){\bf D}
	+2\alpha^2 {\dot {\bf D}},
\end{equation}
with $\nabla\cdot {\bf u}=0$, ${\bf D}=(1/2)(\nabla {\bf u}+\nabla {\bf
u}^{\rm T})$,
$\boldsymbol{\Omega}=(1/2)(\nabla{\bf u}-\nabla{\bf u}^{\rm T})$,
and co-rotational (Jaumann) derivative given by
${\dot {\bf D}}={d}{\bf D}/{dt}+{\bf
D}\boldsymbol{\Omega}-\boldsymbol{\Omega}{\bf D}$.
In this form, one recognizes the constitutive relation for VCHE as a
variant of the rate-dependent incompressible homogeneous fluid of second
grade~\cite{Dunn74},~\cite{Dunn95}, whose viscous dissipation, however,
is modified by the Helmholtz operator $(1-\alpha^2\Delta)$.
There is a tradition at least since Rivlin~\cite{Rivlin57} of modeling
turbulence by using continuum mechanics principles such as objectivity and
material frame indifference. For example, this sort of approach is taken in
deriving Reynolds stress algebraic equation models~\cite{Shih95}.
Rate-dependent closure models of mean turbulence have also been obtained
by the two-scale DIA approach~\cite{Yoshizawa84} and by the renormalization
group method~\cite{Rubinstein90}. Since VCHE describe
mean quantities, we propose to use (\ref{VCHE-eq}), or equivalently
(\ref{VCHE-tnsr}),
as a turbulence closure model and test this ansatz by applying it to
turbulent channel
flow. The corresponding closure for turbulent pipe flows will be treated
elsewhere~\cite{BCP}.

We denote the velocity {\bf u} in (\ref{VCHE-eq}) by ${\bf U}$ and
seek its steady state solutions in the form ${\bf U}=(U(z),0,0)^T$ subject
to the boundary condition $U(\pm{d})=0$ and the symmetry condition
$U(z)=U(-z)$.
In this particular case, the steady VCHE reduces to,
\begin{eqnarray} \label{steady-PCH}
&& -\nu U'' + \nu\alpha^2 U'''' = -\partial_x p \,,
\quad 0 = -\partial_y p\,,
\nonumber \\
&& 0 = -\partial_z \Big( p -\alpha^2 (U')^2\Big)\,.
\end{eqnarray}
That is, $d{\bf v}/dt$ vanishes in~(\ref{VCHE-eq}) and the remaining 
$\alpha^2$ terms modify the pressure and dissipation.
Comparing (\ref{Reynolds-eqns}) and (\ref{steady-PCH}), we
identify counterparts as,
\begin{eqnarray}
\label{Identification}
  &\overline{U}  =  U\,, \quad
 \partial_z \langle wu \rangle = \nu \alpha^2 U^{''''}+p_0\,, \quad
  \partial_z \langle wv \rangle = 0\,,&
 \\
  &\nabla (\overline{P}+\langle w^2 \rangle)
   =  \nabla (p-p_0 x -\alpha^2 (U')^2)\,,&
\nonumber
\end{eqnarray}
for a constant $p_0$.
This identification gives
\begin{equation}  \label{corr-wv+ww}
\hspace{-.3in}\langle wv \rangle (z) = 0,\quad
- \langle wu \rangle (z) = -p_0 z - \nu \alpha^2 U'''(z)\,,
\end{equation}
and leaves $\langle w^2 \rangle$ undetermined up to an arbitrary function
of $z$.
A closure relation for $-\langle wu \rangle$ involving $ U'''(z)$ also
appears in Yoshizawa~\cite{Yoshizawa84}, cf. also equation (8)
of~\cite{Wei Willmarth89}.

The solution of the steady VCHE (\ref{steady-PCH}) in a channel subject to
these boundary and symmetry conditions~\cite{footnote1} is
 \begin{equation}
\label{CH_solution}
  U(z)=a \Big( 1-\frac{\cosh(z/\alpha)}{\cosh(d/\alpha)} \Big)
      +b \Big( 1-\frac{z^2}{ d^2} \Big)
\end{equation}
with constants $a$, $b$. Any time dependent
solution of~(\ref{VCHE-eq}) such that
${\bf u}(z,t)=(U(z,t), 0, 0)^T$ with $U(z,t)=U(-z,t)$ and $U(\pm d,t)=0$
converges exponentially in time to the solution~(\ref{CH_solution})
with the same mean flow and boundary shear stresses \cite{footnote2}.

We consider channel flows for $R>>1$. Our basic assumption is
that $\xi=d/\alpha\to\infty$ as $R\to\infty$.
The solution (\ref{CH_solution}) must satisfy
the stress condition (\ref{stress-cond}),
which imposes 
\begin{equation} \label{strs-1}
\tau_0 = \nu U'(z)|_{z=-d} = \frac{a\nu}{\alpha}{\rm tanh}\,\xi +
\frac{2b\nu}{d}.
\end{equation}
Substituting the definitions,
\begin{eqnarray} \label{def's}
\overline{u}=\frac{1}{2d}\int^d_{-d}U(z)dz
            = a\Big(1-\xi^{-1}{\rm tanh}\,\xi\Big) + \frac{2}{3}b,
 \\
\hbox{and} \quad
R= \frac{\overline{u}d}{\nu} ,
\quad
R_0= \frac{\tau_0^{1/2}d}{\nu} ,
\quad
\theta= \frac{2b\nu}{d\tau_0} ,
\nonumber
\end{eqnarray}
with $0<\theta<1$, into relation (\ref{strs-1}) re-expresses it as
\begin{equation} \label{basic-rel-1}
\hspace{-.3in} R_0^2 =
\Big( R - \frac{\theta R_0^2}{3} \Big)
\Big(\frac{\xi\,{\rm tanh}\,\xi}
{1-\xi^{-1}{\rm tanh}\,\xi}\Big)
+ \theta R_0^2\,.
\end{equation}
Upon setting $\xi=\delta R_0$ and taking $\xi>>1$,
this simplifies to order
$O(1/\xi)$ into the basic relation
\begin{equation} \label{basic-rel-2}
\frac{1-\theta}{\delta} = \frac{R}{R_0}
 - \frac{\theta R_0}{3}\,.
\end{equation}
In terms of the parameters $\theta$ and $\xi$, equations
(\ref{corr-wv+ww}) and (\ref{CH_solution})
imply the Reynolds shear stress,
\begin{equation} \label{wu-profile}
 - \langle wu \rangle (z) = \tau_0(1-\theta)
   \Big[\frac{{\rm sinh}(z/\alpha)}{{\rm sinh}(\xi)} - \frac{z}{d}\Big].
\end{equation}
Thus, the solution (\ref{CH_solution}) implies
$- \langle{wu}\rangle\ge0$ for $-d\le{z}\le0$, as seen
empirically~\cite{Wei Willmarth89}.
In the lower half of the channel,
the symmetric solution (\ref{CH_solution}) may be
expressed in wall units using the notation
$\phi(\eta)=U(z)/u_*$, $\eta=(z+d)/\ell_*$, with
$\ell_*=\nu/u_*=d/R_0$, to order $O(\xi e^{-\xi})$ as
\begin{equation} \label{phi-profile}
\hspace{-.3in}
\phi(\eta)=\frac{1-\theta}{\delta}\Big(1-e^{-\delta\eta}\Big)
	+\theta\eta\Big(1-\frac{\eta}{2R_0}\Big) \,,
\end{equation}
for $0\le\eta\le R_0$.
In this notation, we have $\alpha = d/\xi = \ell_*/\delta$ for the lengthscale
in (\ref{VCHE-eq}). The velocity profile $U(z)$ in (\ref{CH_solution})
has its maximum at the center of the channel $(\eta=R_0)$. At this point
$\phi_{max}\simeq\phi(R_0)$ is given to leading order by
\begin{equation} \label{phi-max}
\phi_{max} = \frac{1-\theta}{\delta} + \frac{\theta R_0}{2}.
\end{equation}
Hence, from (\ref{basic-rel-2}), (\ref{phi-max}) and $\sqrt{2/D}=R/R_0$ we have
\begin{equation} \label{phi-max-R}
\hspace{-.3in}
\theta R_0 = 6\Big(\phi_{max}-\sqrt{\frac{2}{D}}\Big),
\quad
\frac{1-\theta}{\delta} = 3\sqrt{\frac{2}{D}}- 2\phi_{max}.
\end{equation}
Since $0<\theta<1$, relations (\ref{phi-max-R}) imply the inequalities
$3/2 > \phi_{max}\sqrt{{D}/{2}} >1$, and we may write
\begin{equation} \label{t-d-phi-rel} \hspace{-.3in}
\theta R_0 = 3(1-c) \sqrt{\frac{2}{D}}, \quad \hbox{and} \quad
\frac{1-\theta}{\delta} = c \sqrt{\frac{2}{D}} \,,
\end{equation}
by introducing the quantity $c\in(0,1)$ defined in terms of the
velocity profile flatness or midplane velocity ratio
$u_{max}/\overline{u}$ as
\begin{equation} \label{c-u-rel} \hspace{-.4in}
\frac{3-c}{2} = \frac{u_{max}}{\overline{u}}
= \phi_{max}\sqrt{\frac{D}{2}}
\quad \hbox{with} \quad
0<c<1.
\end{equation}
Comparison with data will show that $c$ is nearly independent of
$R$ and bounded away from its endpoint values.
The first equation in (\ref{t-d-phi-rel}) and the basic
relation (\ref{basic-rel-2}) then imply
\begin{equation} \label{theta-eqn}
\theta = \Big(1 + \frac{c\xi}{3(1-c)}\Big)^{-1}= O(\frac{1}{\xi})\,.
\end{equation}
Substituting this into the second equation in (\ref{t-d-phi-rel}) gives,
\begin{equation} \label{R0-R-rel} \hspace{-.3in}
R_0 = c \delta R \Big(1 + \frac{3(1-c)}{c\xi}\Big)
    = c \delta R + O(\frac{1}{\xi})\,.
\end{equation}
Thus, to leading order,
$\delta = c^{-1}R_0/R = c^{-1}\sqrt{{D}/{2}}$ and the
velocity profile (\ref{phi-profile}) is given by
\begin{equation} \label{phi-profile-2} \hspace{-.3in}
\phi(\eta) = \frac{R}{R_0}
\Big[c \Big(1-e^{-\frac{R_0\eta}{cR}}\Big)
+ 3(1-c)\frac{\eta}{R_0}\Big(1-\frac{\eta}{2R_0}\Big)\Big] .
\end{equation}
Thus, for $\xi=d/\alpha >> 1$, the drag $\sqrt{D/2}=R_0/R$ and the
constant $c$ determine the steady velocity profile of VCHE $\phi(\eta)$ at each
$R$.
The lengthscale
$\alpha$ is given to leading order by $\alpha/d=1/(\delta R_0)=2c/(DR)$. For the
Blasius drag law, $D=\lambda{R}^{-1/4}$, $\lambda=const$, this implies
\begin{equation} \label{xa-eqn}
\frac{1}{\xi} = \frac{\alpha}{d}
= \frac{2c}{\lambda} R^{-3/4}
= c\Big(\frac{2}{\lambda}\Big)^{4/7} R_0^{-6/7}\,.
\end{equation}
This scaling in $R$ agrees with the scaling law for the Kolmogorov fluctuation
dissipation length, $\ell_d$. Thus, in this case, $\alpha$ is proportional
to $\ell_d$~\cite{footnote3}. For the Blasius drag law, we also have from~(\ref{t-d-phi-rel})
\begin{eqnarray}
\label{B-imp}
R_0 = \sqrt{\frac{\lambda}{2}}R^{7/8}, \quad
\delta = \frac{\sqrt{\lambda/2}}{c} R^{-1/8}, \nonumber \\
\theta = \frac{3(1-c)}{\lambda/2} R^{-3/4}, \quad
\frac{1-\theta}{\delta} = \frac{c}{\sqrt{\lambda/2}}R^{1/8}.
\end{eqnarray}

\psfig{file=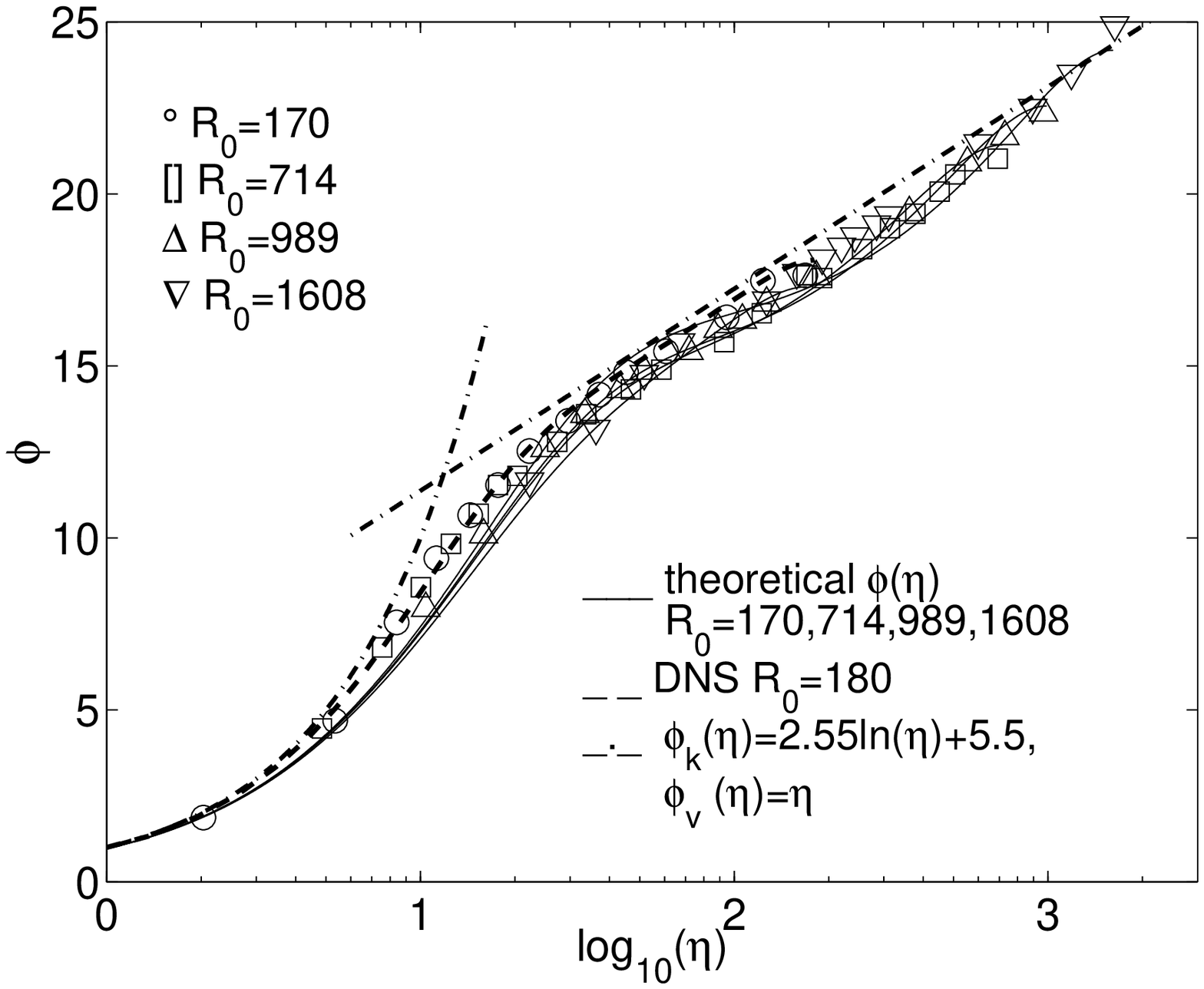,width=220pt}
\noindent
\vskip 5pt
{\small FIG.~1. The mean-velocity profile $\phi$ as a function of $\log(\eta)$.
Experimental data~\cite{Wei Willmarth89},
DNS data~\cite{Kim-Moin-Moser87}. 
$\phi_v$ and $\phi_k$ represent the viscous relation and von
K\'arm\'an logarithm law respectively.}
\smallskip

\psfig{file=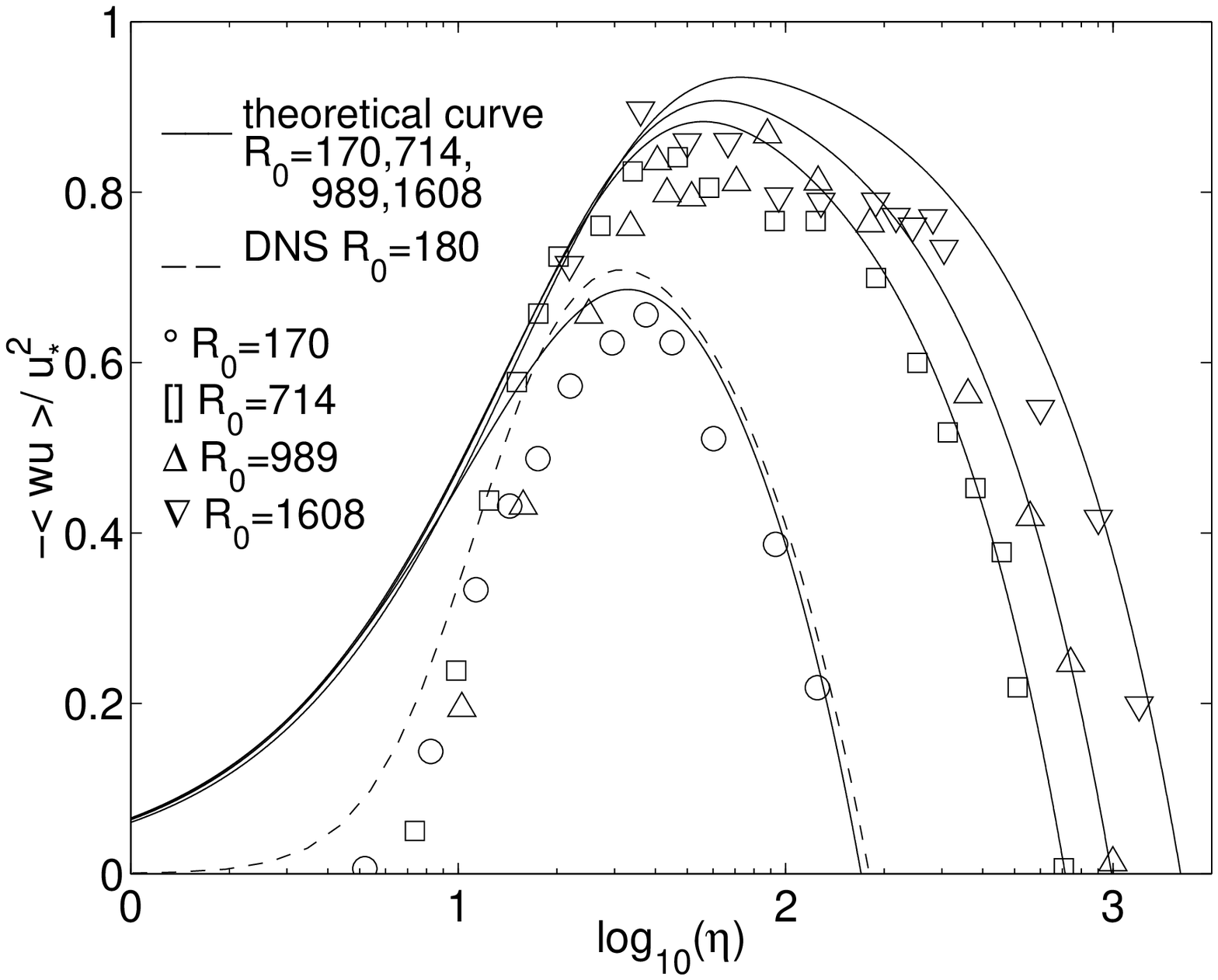,width=220pt}
\noindent
\vskip 5pt
{\small FIG.~2. The Reynolds-stress $-\langle wu\rangle/u_*^2$.
Experimental data~\cite{Wei Willmarth89},
DNS data~\cite{Kim-Moin-Moser87}.}

\medskip
Figures 1--3 compare our theoretical solution~(\ref{phi-profile-2}) 
with experimental
data from~\cite{Wei Willmarth89}, and direct numerical
simulation (DNS) data from~\cite{Kim-Moin-Moser87}.
We assume the Blasius drag law, $D=\lambda R^{-1/4}$, with $\lambda=0.06$.
We vary $c$ slightly with $R$ in order to best fit the
data ($c \in [.728,.77]$).  Note, the curves
can be brought into even better agreement by allowing $\lambda$ to also
vary slightly with~$R$.

\psfig{file=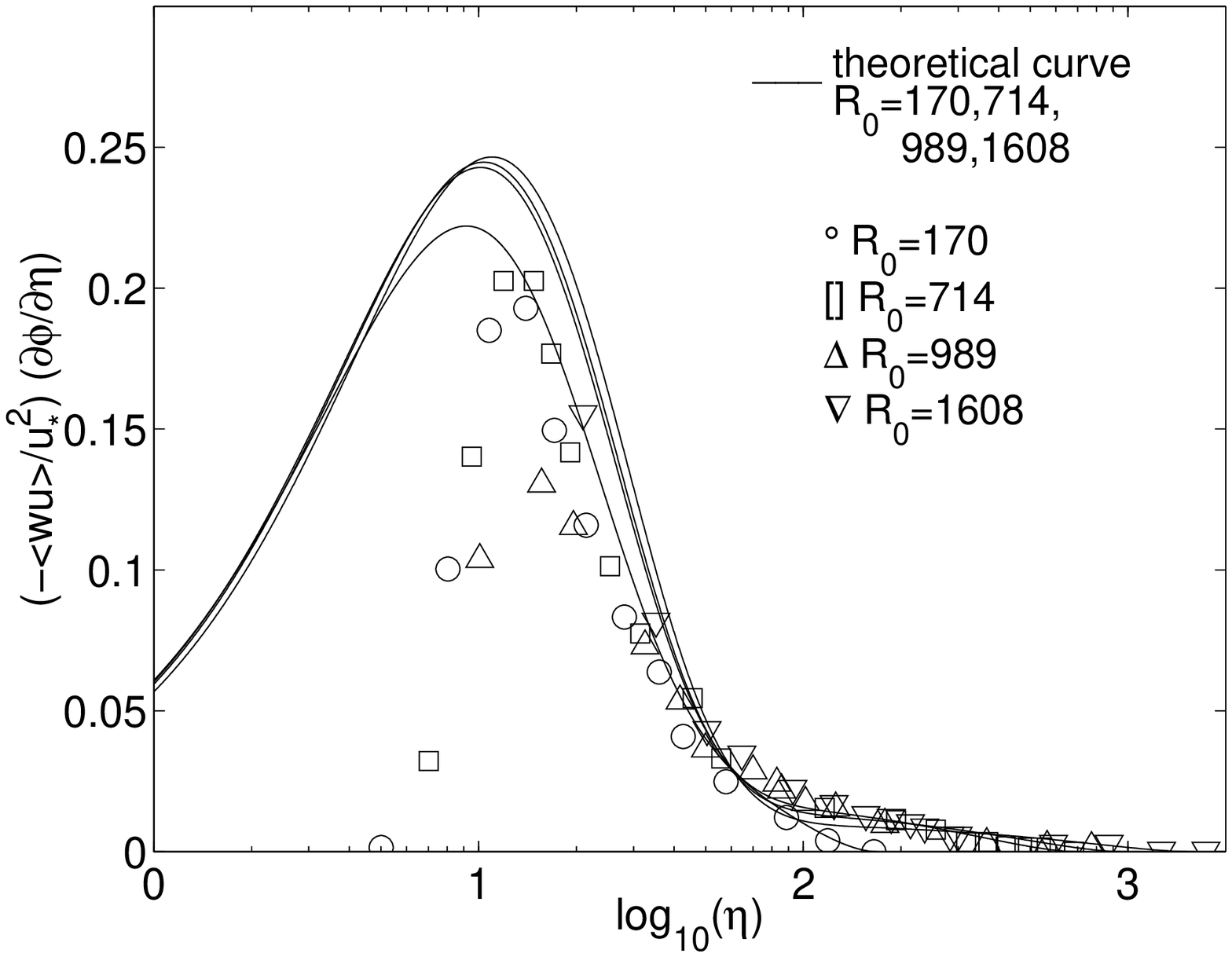,width=220pt}
\noindent
\vskip 5pt
{\small FIG.~3. The turbulent kinetic energy production profile.
Experimental data is from Wei \& Willmarth~\cite{Wei Willmarth89}.}

\bigskip
\psfig{file=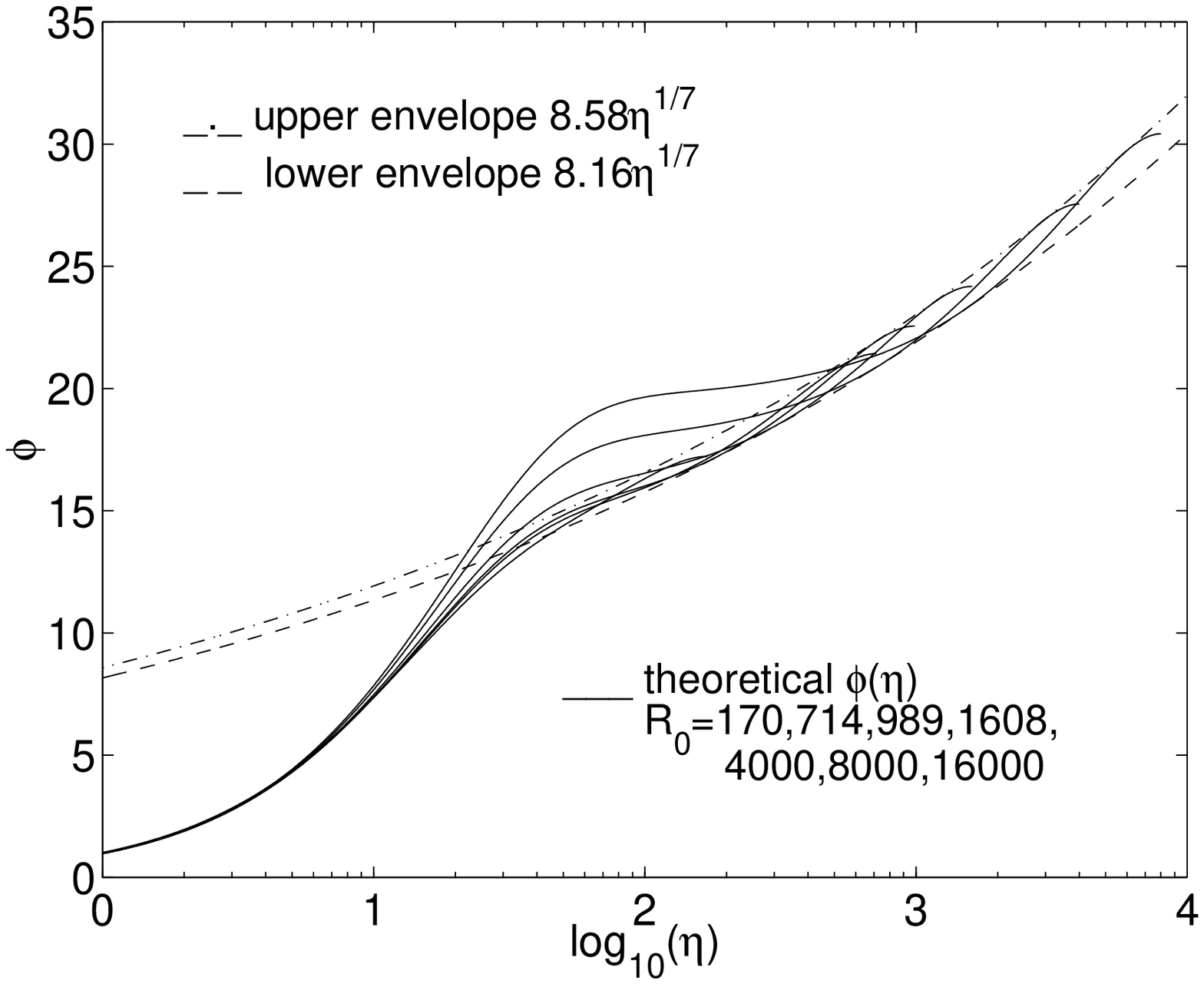,width=220pt}
\noindent
\vskip 5pt
{\small FIG.~4. The upper and lower envelopes of the velocity profile
for the Blasius drag law.}

\bigskip
After the standard Blasius drag law is chosen and $c$ is determined from the
midplane velocity data, no free parameters remain in the model. Figure~2
compares the
theoretical prediction with the measured data for the Reynolds-stress
$-\langle{wu}\rangle/u_*^2$. Figure~3 compares the predicted and measured
turbulent
kinetic energy production profiles. Except for the region nearest the wall,
$\log_{10}(\eta) \le 1$ in Figure~2 and Figure~3, the
theoretical model shows excellent agreement with the data. Using the
fundamental relation (\ref{phi-profile-2}) yields a family of
velocity profiles $\phi(\eta;R)$, plotted in Figure~4 at various values of
$R$ for the Blasius drag law. Remarkably, the upper and
lower envelopes of this family are found analytically to be
$\eta^{1/7}$ power laws. If instead of the Blasius drag law, one uses
the von K\'arm\'an drag law, $\sqrt{2/D}=\lambda_1 \log(R_0)+\lambda_2$, for
constants $\lambda_1$ and $\lambda_2$, then (\ref{phi-profile-2})
yields a family of velocity profiles $\phi(\eta;R)$ whose upper and lower
envelopes are nearly linear in $\log_{10} \eta$ with different slopes.

We are grateful to R.~Kraichnan for constructive
comments and to D.~Cioranescu for pointing out the
relation between the Camassa-Holm equations and
the second-grade fluids.

\end{multicols}


\begin{thebibliography}{99}

\bibitem{Reichhardt38}
  H. Reichhardt,
  Naturwissenschaften \textbf{24/25} (1938), 404.

\bibitem{Eckelmann70}
  H. Eckelmann,
  \textit{Mitteilungen aus dem Max Planck Institut},
  G\"{o}ttingen, no. 48 (1970).

\bibitem{Eckelmann74}
  H. Eckelmann,
  J. Fluid Mech. \textbf{65} (1974), 439--459.

\bibitem{Wei Willmarth89}
  T. Wei and W.W. Willmarth,
  J. Fluid Mech. \textbf{204} (1989), 57--95.

\bibitem{Antonia etal92}
  R.A. Antonia, M. Teitel, J. Kim and L.W.B. Browne,
  J. Fluid Mech. \textbf{236} (1992), 579--605.

\bibitem{Kim-Moin-Moser87}
  J. Kim, P. Moin and R. Moser,
  J. Fluid Mech. \textbf{177} (1987), 133--166.

\bibitem{Hinze75}
	J.O. Hinze,
	\textit{Turbulence},
	Mc-Graw-Hill: New York, 2nd edition (1975).

\bibitem{Townsend}
 A.A. Townsend, {\it The Structure of Turbulent Flow}, Cambridge University
 Press (1967).

\bibitem{Holm et al98}
	D.D. Holm, J.E. Marsden, T.S Ratiu,
	Phys. Rev. Lett., to appear.

\bibitem{CH93}
R. Camassa and	D.D. Holm,
{Phys. Rev. Lett.} {\bf 71} (1993) 1661-1664.

\bibitem{Dunn74}
	J.E. Dunn and R.L. Fosdick,
	Arch. Rat. Mech. Anal. \textbf{56} (1974) 191--252.

\bibitem{Dunn95}
	J.E. Dunn and K.R. Rajagopal,
	Int. J. Engng. Sci. \textbf{33} (1995), 689--729.

\bibitem{Rivlin57}
	R.S. Rivlin,
	Q. Appl. Math. \textbf{15} (1957), 212--215.

\bibitem{Shih95}
	T.H. Shih, J. Zhu, and J.L. Lumley,
	Comput. Methods Appl. Mech. Engrg. \textbf{125} (1995), 287--302.

\bibitem{Yoshizawa84}
	A. Yoshizawa,
	Phys. Fluids \textbf{27} (1984), 1377--1387.

\bibitem{Rubinstein90}
	R. Rubinstein and J.M. Barton,
	Phys. Fluids A, \textbf{2} (1990), 1472--1476.

\bibitem{BCP}
	See, e.g., G.I. Barenblatt, A.J. Chorin and V.M. Prostokishin,
	Appl. Mech. Rev., \textbf{50} (1997), 413--429,
        for a recent survey of pipe flows.

\bibitem{footnote1} With the {\it antisymmetric} steady solution, one may
address
turbulent states of Couette flow by a similar analysis.

\bibitem{footnote2} In fact, the time-dependent VCHE in a periodic box has
unique
classical solutions and a global attractor, whose fractal dimension is
finite and scales
according to Kolmogorov's estimate, $N\sim(L/\ell_d)^3$, where
$\ell_d=(\nu^3/\epsilon)^{1/4}$ is the Kolmogorov dissipation
length, \cite{FHT98}.

\bibitem{FHT98}
  C. Foias, D.D. Holm and E.S. Titi, in preparation.

\bibitem{footnote3}
  Detailed analysis of~(\ref{Identification}) and use of Kolmogorov 
 scaling also yields an
  estimate of the energy-containing macroscale in terms of $\ell_d$ and $d$.
\end{thebibliography}
\end{document}